# Nb$_3$Sn Superconducting Radiofrequency Cavities: a Maturing Technology for Particle Accelerators and Detectors

White Paper for Snowmass 2021 – Topical Group AF7RF

S. Posen (Fermilab), M. Liepe (Cornell U.), G. Eremeev (Fermilab), U. Pudasaini (Jefferson Lab), C.E. Reece (Jefferson Lab)

**Executive summary**

Nb$_3$Sn superconducting radiofrequency (SRF) cavities have substantial potential for enabling new performance capabilities for particle accelerators for high energy physics (HEP), as well as for RF cavity-based detectors for dark matter, gravitational waves, and other quantum sensing applications. Outside of HEP, Nb$_3$Sn SRF cavities can also benefit accelerators for nuclear physics, basic energy sciences, and the industry. The community interest in Nb$_3$Sn SRF is evident from the substantial number of Snowmass 2021 Accelerator Frontier LOIs that discuss Nb$_3$Sn SRF development [1-10].

Nb$_3$Sn has approximately twice as high critical temperature as the standard SRF material, niobium, allowing it to achieve high quality factors even at relatively high temperatures. Nb$_3$Sn cavities have already demonstrated extremely high cryogenic efficiency with promising performance for applications with accelerating gradients ~10 MV/m. Small scale accelerators based on Nb$_3$Sn are now moving towards the prototyping stage. Current Nb$_3$Sn cavities show gradient limitation significantly below niobium cavities. The best Nb$_3$Sn cavities so far can reach slightly over ~20 MV/m, but the best niobium cavities can reach ~50 MV/m. However, experiments suggest that even high performing Nb$_3$Sn cavities are limited by surface defects, and improved surfaces are expected to lead to improved gradients. The superheating field of Nb$_3$Sn would correspond to a maximum accelerating gradient of ~100 MV/m, well above the corresponding value for niobium, which is very close to the currently observed limit of ~50 MV/m. The small superconducting coherence length of Nb$_3$Sn ~3-4 nm may mean that RF surfaces need to be significantly smoother than what is currently achieved to reach the full potential. Thinner films may also help to stabilize the Nb$_3$Sn layer against defects. Very limited research has been performed so far on Nb$_3$Sn compared to niobium, for which advanced surface preparation techniques have been developed. Additional investment in Nb$_3$Sn SRF research is needed to help this material reach much closer to its full potential. Nb$_3$Sn R&D efforts are funding limited, but even with relatively small efforts, there have been significant advances in Nb$_3$Sn performance over the last years.

High efficiency, medium gradient applications for Nb$_3$Sn include high duty factor accelerators, including e+e- circular colliders and intensity frontier front-end machines. High efficiency, high gradient applications include a future linear collider, such as an energy upgrade to the ILC. Dark sector search applications include RF cavities that achieve high quality factors in multi-tesla fields for axion horoscopes. Industrial applications include wastewater treatment, medical isotope production, and ultrafast electron diffraction.

Support for Nb$_3$Sn research will be directed towards: continuing to improve understanding of the influence of film microstructure on maximum gradient; improving the coating process to reduce gradient-limiting defects; new coating approaches including multilayers; and practical demonstrations and problems solving of new engineering requirements.

**Introduction**

Superconducting radiofrequency (SRF) cavities are extremely efficient devices for generating large electromagnetic fields, which often makes them the technology of choice for transferring energy to beams in modern particle accelerators. Major high energy physics facilities that are based on SRF accelerators include LBNF/DUNE [11], LHC [12], HL-LHC [13], and EIC [14], as well as the proposed next generation Higgs factories ILC [15], FCC-ee [16], and CepC [17]. Normal conducting (NCRF) and SRF cavities can both reach accelerating gradients on the order of 10s of MV/m. The key advantage that superconducting cavities have is their high quality factor $Q_0$, which gives them orders of magnitude smaller heat dissipation, allowing them to operate with high duty factor at high fields (e.g. constantly running vs requiring short pulses) and to greatly reduce the amount of overhead from RF power supplies that would be dissipated in the cavity walls.

SRF cavities have been around for over 50 years [18], and all SRF accelerators in use today use niobium as the material in the RF surface. Niobium has been the material of choice because it has good superconducting properties (e.g. high critical temperature ~9.2 K), and, as an element, is easy to fabricate with good stoichiometric uniformity over a large ~1 m$^2$ surface. Over years of development, new cavity treatments have led to continued improvement in Nb cavity performance. For example, the maximum accelerating gradient of Nb cavities has reached as high as ~50 MV/m [19, 20], which is very close to the predicted ultimate limit set by the superheating field of the superconductor [21, 22, 23].

While research continues on improving niobium cavity performance, research effort is also being dedicated in parallel to next-generation SRF cavity materials which have the potential to significantly outperform Nb and thus replace Nb as the prime SRF material. The current most promising and most advanced next-generation material is Nb$_3$Sn [24]. We are therefore expressing here a strong recommendation for increasing support of Nb$_3$Sn SRF research and technology development.

**Medium term motivation: High Q at high T**

Nb$_3$Sn has a critical temperature of ~18 K, about twice that of niobium, allowing it to achieve a high $Q_0 > 10^{10}$ at ~2x higher operating temperatures than Nb. Changing the operating temperature from typical 2.0 K for Nb to 4.4 K for Nb$_3$Sn while maintaining intrinsic quality factors in the $10^{10}$ to $10^{11}$ range would reduce energy consumption and thus cryogenic operating costs by as much as an order of magnitude, and would substantially decrease infrastructure costs for the cryogenic plant.

This would be a considerable advantage for high duty factor HEP accelerators. This includes circular electron positron colliders, such as FCC-ee [16]. The proposed e+e- circular colliders FCC-ee/CePC would require operation of hundreds of SRF cavities (e.g, at 400 and 800 MHz) for a total RF voltage of multiple GeV; above 10 GeV in ttbar running mode [16]. Employing Nb$_3$Sn based SRF cavities (e.g, Nb$_3$Sn @ 800 MHz @ 4.5K @ 20 MV/m) instead of niobium would allow a substantial reduction in the size and operating cost of the required cryogenic plant, and would make it possible to reduce the overall power consumption of these accelerators considerably.

Another high energy physics application could be for high duty factor linear accelerators, which are proposed for several future directions. This includes as a driver for future multi-MW neutrino experiments at Fermilab for LBNF/DUNE (an extension to PIP-II, including a possible booster replacement [25]) as well as a driver for a muon collider [26]. It also includes CW electron accelerators for dark matter searches [28, 29].

For a linear collider (ILC) Higgs Factory and Top Factory upgrade, high-Q, high-temperature $Nb_3Sn$ could enable increasing the RF pulse length as well as the repetition rate of the pulses, thereby greatly increasing luminosity.

**Long term motivation: Potential for high gradient operation**

$Nb_3Sn$ also has a predicted superheating field that is twice as high as niobium [23], which could allow for a similar increase in the maximum accelerating field, up to 100 MV/m. If this could be realized practically, it would be an enormous advantage for high energy linac applications such as an energy upgrade for the ILC to multi-TeV (see also separate LOI on ILC high-energy upgrade [30]).

For some schemes of proton drivers for neutrino and muon based physics, a short duty factor pulsed linear accelerator is called for. High accelerating gradient could help to reduce linac length and cost in this case.

Exceeding the current ~20 MV/m of $Nb_3Sn$ cavities would also greatly benefit the high duty factor HEP accelerator discussed above. The small wall dissipation (high $Q_0$) of $Nb_3Sn$ will allow operating high duty factor SRF cavities at substantially higher optimal field gradients than standard niobium cavities, thereby reducing the number of SRF cavities required, and in turn reducing the impedance of the accelerator.

**Additional motivation: Dark sector searches**

SRF cavities are also being explored not as a means of accelerating particles but as a means for detecting them in the next generation of dark sector searches [31-34]. $Nb_3Sn$ could have a distinct advantage for these searches due to its ability to remain superconducting in large magnetic fields, which would be important for example for axion haloscopes. Initial results have been promising, showing $Q_0$ values of ~$5 \times 10^5$ at 6 T and 4.2 K [35]. Successfully implementing cavities with these $Q_0$ values (while keeping other parameters the same) in an axion search would significantly improve sensitivity and scan rate compared to typical copper cavities.

**Motivation Outside of HEP**

$Nb_3Sn$ cavities have a very high quality factor even above 4 K. They can be cooled with cryocoolers for cavity cooling, which significantly cuts capital and installation cost, enabling compact and potentially even mobile applications. For small-scale applications, $Nb_3Sn$ SRF opens up the possibility for turn-key operation with cryocoolers instead of complex liquid helium cryogenic plants. This greatly reduces the footprint of the system, and substantially simplifies operation and maintenance.

There are expected to be applications in nuclear physics facilities for small $Nb_3Sn$ cryomodules, especially cryocooler-based modules for a small number of cavities that are in an isolated location not near an existing cryogenic distribution system, for example for isotope separation. A specific example of an application already underway is at JLab, for the upgraded injector test facility (UITF), where a quarter cryomodule can be used to accelerate an electron beam up to 10 MeV. The facility can use a cryomodule with $Nb_3Sn$-coated cavities to run low-energy nuclear physics experiments. Installing a 4 K capable cryounit at the front of CEBAF would allow year-round injector development operations, while the rest of the accelerator is down for servicing. One can dream about running the facility using cryocoolers at 4 K with $Nb_3Sn$ cavities instead with cryogenic at 2 K. Developing an $Nb_3Sn$ quarter module suitable to install and test into the UITF with $Nb_3Sn$ cavities is in progress.

In a similar way, $Nb_3Sn$ cavity units could further enable and greatly simplify widespread use of SRF technology in light-source storage rings (e.g., a cryocooler based single-cell $Nb_3Sn$ 500 MHz cavity cryomodule), FELs, and compact accelerators for biosciences and material science, e.g., using ultrafast electron diffraction (UED).

There have been tests of $Nb_3Sn$ cavities operating in conduction cooling setups as demonstrations for industrial accelerator applications at Fermilab, JLab, and Cornell [36-38]. The Fermilab demonstration was a $Nb_3Sn$ coating on an all-niobium 650 MHz single cell cavity with welded conduction cooling rings. JLab tested a $Nb_3Sn$ coating on a 1.5 GHz single cell cavity, with a copper layer deposited on the outer side by cold-spray (~of 76 μm) followed by copper plating (>= 5 mm) for conduction cooling. The Cornell demonstration was on a $Nb_3Sn$-coated 2.6 GHz single cell Nb cavity with cooling links clamped to its beamtubes.

There are also ongoing efforts to build demonstration $Nb_3Sn$ cryomodules for industrial applications. Detailed plans have been published for a medium energy, high average power superconducting e-beam accelerator for environmental applications by researchers at Fermilab [39] and a a cw, low-energy, high-power superconducting linac for environmental applications by researchers at JLab [40]. A conduction-cooled $Nb_3Sn$ SRF cryomodule housing a single-cell 1.3 GHz cavity capable of 100 mA beam operation is currently under development at Cornell University. Euclid Techlabs is leading an effort to build a $Nb_3Sn$ cryomodule for ultrafast electron microscopy applications [41].

**Current landscape for $Nb_3Sn$ R&D**

In the U.S, the Department of Energy started funding $Nb_3Sn$ R&D initially at Cornell University, followed by programs at Jefferson Lab and Fermilab [24]. Stimulated by the $Nb_3Sn$ SRF progress at these laboratories with first proof-of-principle demonstrations of superior performance, worldwide interest in $Nb_3Sn$ SRF has greatly increased recently, and new $Nb_3Sn$ R&D efforts have started at labs and in industry, *e.g.* at CERN, IMP, ULVAC/KEK, NHMFL/Florida State University/University of Texas–Arlington, Peking University, STFC, ODU, and Ultramet [42-49].

$Nb_3Sn$ cavity performance has not reached its ultimate performance potential discussed above yet, but it has been making substantial progress over the past years. Using as a metric the maximum accelerating gradient with $Q_0>10^{10}$ at 4.4 K, cavities have increased from ~5 MV/m in the 1990s [50] to ~13 MV/m in 2014 [51], to ~18 MV/m in 2015 [52], to ~24 MV/m in 2020 [53]. This has come with corresponding improvements in understanding of the materials science and fabrication methods for the $Nb_3Sn$ coatings [54-61].

Current state-of-the-art $Nb_3Sn$ cavities show gradient limitation significantly below niobium cavities. The best $Nb_3Sn$ cavities so far can reach slightly over ~20 MV/m, but the best niobium cavities can reach ~50 MV/m. However, experiments suggest that even high performing $Nb_3Sn$ cavities are limited by surface defects [62,63], and improved surfaces are expected to lead to improved gradients. The superheating field of $Nb_3Sn$ would correspond to a maximum accelerating gradient of ~100 MV/m, well above the corresponding value for niobium, which is very close to the currently observed limit of ~50 MV/m. $Nb_3Sn$ is expected to be more sensitive to surface defects than niobium because of its relatively short coherence length ~3-4 nm, approximately an order of magnitude smaller than niobium, depending on the niobium surface treatment. On the other hand, $Nb_3Sn$ R&D on surface treatments after coating is still relatively primitive compared to niobium. Coated surfaces are relatively rough on relevant length scales, and attempts to smooth surfaces using techniques developed for niobium have so far resulted in other issues such as residues and performance degradation [64,65].

Compared to niobium, relatively little effort has so far been invested in surface processing of $Nb_3Sn$. New efforts are underway to smoothen $Nb_3Sn$ surfaces and also make the coatings thinner, which could help to thermally stabilize any defects that are present by reducing thermal impedance. There are promising directions, including electropolishing, oxypolishing, and mechanical polishing, as well as new deposition methods to try to create inherently smoother films.

According to the concept proposed by A. Gurevich [66], it is possible to significantly enhance the RF breakdown of the magnetic field of SRF cavity by engineering multilayer thin film of alternating insulating layers of superconducting layers of thickness smaller than the London penetration depth. $Nb_3Sn$ is a potential thin film to be used in such multilayer structures. For example, it has been estimated that the optimized thickness of the ideal $Nb_3Sn$ and insulating layer can result in a maximum field of 400 mT compared to 200 mT [67]. Despite an attractive approach, it is significantly challenging to deposit defect-free layers of superconductors and insulators. Recent progress of novel deposition techniques such as energetic ion vacuum deposition, atomic layer deposition, etc. allows producing well-engineered single superconducting layer (e.g., Nb on Cu) or multilayered superconductor-insulator-superconductor structures. Research is progressing toward developing SRF cavities deposited with high-quality Nb and multilayer structures based on $Nb_3Sn$ using energetic ion deposition to avoid high-temperature treatment and result in high-quality material layers.

Superconducting radiofrequency technology relies on niobium and its superconducting properties to achieve the gradient and quality factor specifications of accelerator projects. After decades of research, niobium material has been pushed close to the superheating field of what is expected from an ultrapure niobium material. To overcome this limit and to achieve higher RF fields several ideas have been put forth on how to modify the surface and material within the penetration depth. As an example, a paper by Ngampruetikorn and Sauls [68] discusses how inhomogeneous surface disorder can increase the superheating field above that for a clean surface. The authors argue that an inhomogeneous disorder, e.g., an impurity diffusion layer, can increase the superheating field above that in the clean limit by distributing the screening currents into cleaner regions with larger critical currents, hence, achieve a higher surface superheating field. While these ideas typically analyze niobium, the underlying ideas apply not only to niobium, but also other superconductors such as $Nb_3Sn$. A long-term R&D in $Nb_3Sn$ can exploit, verify this.

**Technology Development**

The potential for $Nb_3Sn$ material requires further focused R&D to enable even higher accelerating gradients and quality factors. Some of the development challenges for accelerating modules become evident only during engineering and technical developments of operational accelerating modules. One of the challenges that has not been evident during the superconducting material development phase is the mechanical deformation sensitivity of $Nb_3Sn$ films coated on niobium in the vapor diffusion process. This sensitivity, which was identified during assembly of accelerating structures into the accelerating module[69], limits the amount of tuning that can be applied to a coated cavity and may require different tuning schemes to tune the frequency of accelerating structures to the accelerator frequency. Further efforts are needed at both the scientific and technical level to identify and resolve such issues as discussed in the next section. If funded, the synergy of such efforts will enable timely deployment of the latest advances in the fundamental development in the material to the practical units.

Besides elliptical cavities, $Nb_3Sn$ has the potential to be used in cavities of complicated geometries such as twin axis cavities, half coaxial resonators, etc., to simplify their application in various SRF accelerator applications. Because of advanced geometry with hard-to-reach areas, conformal thin film deposition uniformly inside these cavities is challenging. After the vapor deposition of $Nb_3Sn$ at a higher temperature, these cavities are prone to mechanical deformation. At Jefferson Lab, a twin axis cavity was coated, but failed mechanically after the coating on its way to RF testing. Additional engineering effort is needed to design the coating process and preserve the mechanical integrity of the twin-axis cavity after coating until the application.

Operating of $Nb_3Sn$ with cryocoolers for compact, stand-alone applications calls for development and optimization of effective conduction cooling of SRF cavities. Cryomodule concepts will have to be developed that are optimized for this novel cavity cooling method, with low static heat loads, but also the potential for significant simplifications as compared to cryogen based SRF modules. High beam power operation of such cryocooler conduction-cooled $Nb_3Sn$ cavity modules will need high-power RF input couplers with low static and dynamic heat loads.

## Recommendation for continued investment in $Nb_3Sn$

Superconducting RF is a key technology for future HEP accelerators. With continued investment into $Nb_3Sn$ SRF cavity R&D, next-generation cavities based on $Nb_3Sn$ will become a reality with performance specifications highly beneficial for HEP applications, enabling higher luminosity, higher energy, and facilitating energy sustainable science.

Continued investment in $Nb_3Sn$ cavity R&D will provide excellent training opportunities for graduate students in SRF science and technology. SRF experts are a critical need for the future of the U.S. national accelerator labs, and soon will be for industry providing and employing SRF technology based on $Nb_3Sn$. The U.S. DOE funded $Nb_3Sn$ research program has already graduated several scientists in SRF with broad, hands-on expertise. As such, continued $Nb_3Sn$ research will not only provide next-generation technology needed for future accelerator facilities, but also the future leaders that will help develop - and ultimately bring to reality - the accelerators that will drive the HEP research program as well as the research program in a wide range of other sciences over the next decades.

Funding should address both fundamental developments in superconducting $Nb_3Sn$ material, supporting material deposition techniques, surface and RF research, and should also address technical challenges, supporting system-level developments (e.g., tuning systems). Practical accelerating modules, exploiting the state-of-the-art $Nb_3Sn$ technology, require both developments in material performance and auxiliary component development to operate in this novel mode. Funding to both fundamental and engineering development needs to be expanded and provided to deliver practical useful accelerator units for the future accelerators.

## Conclusions

The unique advantages of $Nb_3Sn$ cavities make them an exciting prospect for future HEP experiments. They already achieve high $Q_0$ at 4.4 K at accelerating gradients that are useful for high duty factor applications, including first demonstrations in large, accelerator-scale structures [53, 56]. Efforts are underway, including support from sources outside of typical HEP basic research funding, including

SBIRs [41], the National Nuclear Security Administration (NNSA), and the US Army Engineer Research and Development Center (ERDC) to develop first practical small-scale accelerators based on $Nb_3Sn$. Success in first applications would make $Nb_3Sn$ attractive for high duty factor linear accelerators, with possible applications in high energy physics that include circular colliders and high intensity drivers for neutrino- and muon-based physics. With continued materials development progress, $Nb_3Sn$ cavities have the potential to further reduce cryogenic losses and also to eventually outperform current state-of-the-art niobium in energy gain by a significant margin for high energy linear accelerator applications. Initial investigations show that $Nb_3Sn$ cavities are promising for dark sector searches requiring a high magnetic field [35], showing high $Q_0$ in multi-tesla fields, and a corresponding increase in detector sensitivity and scan rate. Continued development could help to increase $Q_0$ even further in this high magnetic field regime. We ask the Snowmass community to strongly endorse continued investment into $Nb_3Sn$ SRF cavity R&D so that the potential of this material can be realized and exploited for high energy physics, nuclear physics, basic energy sciences, and industrial accelerators.

**References**
[1] C. Thangaraj and T. Kroc, "Compact Superconducting RF Accelerators," Snowmass 2021 LoI, Accelerator Frontier #070.
[2] H. Padamsee, "Perspectives on Superconducting Linear Colliders (ILC) to the Next Century Part B: ILC Energy Upgrades to 3 TeV and Beyond," Snowmass 2021 LoI, Accelerator Frontier #075.
[3] G. Bisoffi and L. Rossi, "INFN Position Paper for Snowmass'21 on Accelerators," Snowmass 2021 LoI, Accelerator Frontier #100.
[4] M. Seidel (on behalf of the ICFA panel for sustainable accelerators and colliders), "Fostering the development of energy efficient and sustainable technologies and concepts for accelerator driven research infrastructures," Snowmass 2021 LoI, Accelerator Frontier #118.
[5] M. Benedikt, E. Jensen, R. Rimmer, J. Seryi, K. Smith, F. Willeke, and F. Zimmermann, "Superconducting RF for CW Synchrotrons and ERLs," Snowmass 2021 LoI, Accelerator Frontier #152.
[6] U. Pudasaini, C.E. Reece, R.A. Rimmer, A-M. Valente-Feliciano, E. Barzi, and G. Eremeev, "Next Generation SRF accelerators based on Nb3Sn," Snowmass 2021 LoI, Accelerator Frontier #172.
[7] A.-M. Valente-Feliciano, C. Antoine, S. Anlage, J. Delayen, F. Gerigk, A. Gurevich, T. Junginger, S. Keckert, G. Keppel, J. Knobloch, T. Kubo, O. Kugeler, D. Manos, C. Pira, U. Pudasaini, C.E. Reece, R.A. Rimmer, G.J. Rosaz, T. Saeki, R. Vaglio, R. Valizadeh, W. Venturini Delsolaro, P. B. Welander, and M. Wenskat, "Next-Generation Superconducting RF Technology based on Advanced Thin Film Technologies and Innovative Materials for Accelerator Enhanced Performance & Energy Reach," Snowmass 2021 LoI, Accelerator Frontier #205.
[8] S. Posen, D. Bafia, S. Balachandran, S. Belomestnykh, M. Bertucci, A. Burrill, A. Cano, M. Checchin, G. Ciovati, L.D. Cooley, G. Dalla Lana Semione, J. Delayen, G. Eremeev, F. Furuta, F. Gerigk, B. Giaccone, D. Gonnella, A. Grassellino, A. Gurevich, W. Hillert, M. Iavarone, J. Knobloch, T. Kubo, W.-K. Kwok, R. Laxdal, P.J. Lee, M. Liepe, M. Martinello, O.S. Melnychuk, A. Nassiri, A. Netepenko, H. Padamsee, C. Pagani, A. Palczewski, R. Paparella, U. Pudasaini, C.E. Reece, D. Reschke, A. Romanenko, M. Ross, K. Saito, J. Sauls, D.N. Seidman, T. Spina, N. Solyak, Z. Sung, K. Umemori, A.-M. Valente-Feliciano, W. Venturini Delsolaro, N. Walker, H. Weise, U. Welp, M. Wenskat, G. Wu, X. X. Xi, V. Yakovlev, A. Yamamoto, J. Zasadzinski, "Key Directions for Research and Development of Superconducting Radiofrequency (SRF) Cavities," Snowmass 2021 LoI, Accelerator Frontier #206.

[9] S. Balachandran, P.J Lee, W. Withanage, K. Solanki, F. Pourboghrat, L.D Cooley, "The necessity of a basic materials research community for the accelerated development of SRF materials," Snowmass 2021 LoI, Accelerator Frontier #229.
[10] E. Barzi, B. Barish, R. A. Rimmer, A.-M. Valente-Feliciano, B. Barletta, M. Ross, P. B. Welander, L. Alff, N. Karabas, M. Major, J.P. Palakkal, S. Petzold, N. Pietralla, N. Schäfer, A. Kikuchi, H. Hayano, H. Ito, E. Kako, K. Umemori, A. Yamamoto, "An Impartial Perspective for Superconducting Nb3Sn coated Copper RF Cavities for Future Linear Accelerators," Snowmass 2021 LoI, Accelerator Frontier #248.
[11] B. Abi et al., "Deep Underground Neutrino Experiment (DUNE) Far Detector Technical Design Report," arXiv:2002.03005 (2020).
[12] O.S. Brüning et al., "LHC Design Report," Report number CERN-2004-003-V-1 (2004).
[13] G. Apollinari et al. (eds.), "High-Luminosity Large Hadron Collider (HL-LHC): Technical Design Report V.01," CERN Yellow Reports: Monographs CERN-2017-007-M (2017).
[14] F. Willeke et al., "Electron Ion Collider Conceptual Design Report 2021," BNL-221006-2021-FORE (2021).
[15] T. Behnke et al. (eds.), "The International Linear Collider Technical Design Report," arxiv:1306.6327 (2013).
[16] A. Abada et al. "FCC-ee: The Lepton Collider: Future Circular Collider Conceptual Design Report," *Eur. Phys. J. Special Topics* 228, 261–623 (2019).
[17] da Costa, J.G. et al. (eds.) "CEPC Conceptual Design Report," arxiv: 1811.10545 (2018).
[18] H. Padamsee, "50 years of success for SRF accelerators—a review," *Supercond. Sci. Technol.* 30 053003 (2017).
[19] D. Bafia et al., "Gradients of 50 MV/m in TESLA shaped cavities via modified low temperature bake," Proc. 19th Int. Conf. on RF Superconductivity, Dresden, Germany, TUP061 (2019).
[20] R.L. Geng et al., "High Gradient Studies for ILC with Single-Cell Re-Entrant Shape and Elliptical Shape Cavities made of Fine-Grain and Large-Grain Niobium," Proc. PAC07, Albuquerque, New Mexico, USA, WEPMS006 (2007).
[21] C.P. Bean and J.D. Livingston, "Surface barrier in type-II superconductors," *Phys. Rev. Lett.* 12.1 14 (1964).
[22] J. Matricon and D. Saint-James "Superheating fields in superconductors," *Phys. Lett. A* 24 241–2 (1967).
[23] G. Catelani and J.P. Sethna "Temperature dependence of the superheating field for superconductors in the high-κ London limit," *Phys. Rev. B* 78 224509 (2008).
[24] S. Posen and D.L. Hall, "Nb3Sn superconducting radiofrequency cavities: fabrication, results, properties, and prospects," *Supercond. Sci. Technol.* 30 033004 (2017).
[25] M. Checchin, D. Johnson, D. Neuffer, S. Posen, N. Solyak, V. Yakovlev, "Letter of Interest in a Linac-based 8 GeV Accelerator for Booster Replacement and other Applications" Snowmass 2021 LoI, Accelerator Frontier #072.
[26] International Muon Collider Collaboration, "Muon Collider Accelerator Facility," Snowmass 2021 LoI, Accelerator Frontier #102.
[28] P. Achenbach, M. Battaglieri, A. Bianconi, M. Bondì, A. Celentano, G. Costantini, P. Cole, R. De Vita, A. D'Angelo, A. Denig, L. Doria, M. De Napoli, G. Krnjaic, A. Italiano, H-S. Jo, L. Lanza, M. Leali, L. Marsicano, V. Mascagna, H. Merkel, N. Randazzo, E. Santopinto, E. Smith, D. Snowden-Ifft, S.

Stepanyan, M. Ungaro, L. Venturelli, and M. Wood, "Beam Dump eXperiments with electron beams,"Snowmass 2021 LoI, Rare Processes and Precision Measurements Frontier #76.

[29] Y. Nosochkov, T. Beukers, A. Fry, C. Hast, T. W. Markiewicz, T. K. Nelson, N. Phinney, T. O. Raubenheimer, P. C. Schuster, N. Toro, "Dark sector experiments at LCLS-II (DASEL) accelerator design," *Proceedings of the 8th International Particle Accelerator Conference (IPAC 2017)*, Copenhagen, Denmark, May 14 – 19 (2017).

[30] "Perspectives on Superconducting Linear Colliders (ILC) to the Next Century Part B: ILC Energy Upgrades to 3 TeV and Beyond," Snowmass 2021 LoI, Accelerator Frontier #75.

[31] A. Sonnenschein, "Axion Dark Matter eXperiment (ADMX) 2-4 GHz," Snowmass 2021 LoI, Cosmic Frontier #93.

[32] T. Braine et al., "Extended Search for the Invisible Axion with the Axion Dark Matter Experiment," *Phys. Rev. Lett.*, 124, 101303 (2020)

[33] M. Tobar et al., "Low mass, UP-conversion Loop Oscillator Axion Detector using a Microwave Cavity (UPLOAD-MC)," Snowmass 2021 LoI.

[34] A. Grassellino et al., "Dark SRF – experiment," presented to the Fermilab Physics Advisory Committee,
https://indico.fnal.gov/event/19433/contributions/52137/attachments/32415/39710/DarkSRF.pdf

[35] S. Posen, M. Checchin, O.S. Melnychuk, T. Ring, I. Gonin, "Measurement of high quality factor superconducting cavities in tesla-scale magnetic fields for dark matter searches," Arxiv:2201.10733 (2022).

[36] R.C. Dhuley et al, *Supercond. Sci. Technol.* 33 06LT01 (2020).

[37] G. Ciovati et al, *Supercond. Sci. Technol.* 33 07LT01 (2020).

[38] N. Stilin et al, arXiv:2002.11755v1 (2020).

[39] R.C. Dhuley, I. Gonin, K. Zeller, R. Kostin, S. Kazakov, B. Coriton, T. Khabiboulline, A. Sukhanov, V. Yakovlev, A. Saini, N. Solyak, A. Sauers, J.C.T. Thangaraj, "Design of a medium energy, high average power superconducting e-beam accelerator for environmental applications," arXiv:2112.09233 (2021).

[40] G. Ciovati, J. Anderson, B. Coriton, J. Guo, F. Hannon, L. Holland, M. LeSher, F. Marhauser, J. Rathke, R. Rimmer, T. Schultheiss, and V. Vylet, "Design of a cw, low-energy, high-power superconducting linac for environmental applications," *Phys. Rev. Accel. Beams* 21, 091601 (2018).

[41] A. Liu, R. Kostin, C. Jing, P. Avrakhov, "Optimization of an SRF gun design for UEM applications," Proceedings of the North American Particle Accelerator Conference, Lansing, MI, USA (2019).

[42] E.A. Ilyina et al., "Development of sputtered Nb3Sn films on copper substrates for superconducting radiofrequency applications," 2019 Supercond. Sci. Technol. **32** 035002 (2019).

[43] Z. Yang et al., "Development of Nb3Sn Cavity Coating at IMP," Proc. of SRF'19, Dresden, Germany (2019).

[44] R. Ito et al., Nb3Sn Thin Film Coating Method for Superconducting Multilayered Structure," Proc. of SRF'19, Dresden, Germany (2019).

[45] C-U. Kim et al., "Exploration of Electrochemical Processes for Creating Nb3Sn Thin Films via Bronze Route," CAARI 2018 Dallas, Texas, United States August 12- August 17 (2018).

[46] L. Ziao et al., "The Study of Deposition Method of Nb3Sn Film on Cu Substrate," Proc. of SRF'19, Dresden, Germany (2019).

[47] R. Valizadeh at al., "PCD Deposition of Nb3Sn Thin Film of Copper Substrate from an Alloy Nb3Sn Target," Proc. of IPAC'19, Melbourne, Australia (2019).

[48] N. Sayeed et al., "Deposition of Nb3Sn Films by Multilayer Sequential Sputtering for SRF Cavity Application," Proc. of SRF'19, Dresden, Germany (2019).
[49] M. Ge at al., "CVD Coated Copper Substrate SRF Cavity Research at Cornell University," Proc. of SRF'19, Dresden, Germany (2019).
[50] G. Müller, et al. "Status and Prospects of Nb3Sn cavities for superconducting linacs" *Proc. Workshop on Thin Film Coating Methods for Superconducting Accelerating Cavities* TESLA Report 2000-15 (2000).
[51] S. Posen and M. Liepe, "Advances in development of Nb3Sn superconducting radio-frequency cavities," *Phys. Rev. ST Accel. Beams* 17, 112001 (2014).
[52] S. Posen, D.L. Hall, and M. Liepe, "Proof-of-principle demonstration of Nb3Sn superconducting radiofrequency cavities for high Q0 applications," *Appl. Phys. Lett.* 106, 082601 (2015).
[53] S. Posen et al., "Advances in Nb3Sn superconducting radiofrequency cavities towards first practical accelerator applications," *Supercond. Sci. Technol.* 34 025007 (2021).
[54] J. Lee et al., "Grain-boundary structure and segregation in Nb3Sn coatings on Nb for high-performance superconducting radiofrequency cavity applications," *Acta Materialia* 188 pp. 155-165 (2020).
[55] J. Lee et al., "Atomic-scale analyses of Nb3Sn on Nb prepared by vapor diffusion for superconducting radiofrequency cavity applications: A correlative study," *Supercond. Sci. Technol.* 32 024001 (2019).
[56] G. Eremeev et al., "Nb3Sn multicell cavity coating system at Jefferson Lab," *Rev. Sci. Instrum.* 91, 073911 (2020).
[57] U. Pudasaini et al., "Initial growth of tin on niobium for vapor diffusion coating of Nb3Sn" *Supercond. Sci. Technol.* 32 045008 (2019).
[58] U. Pudasaini et al., "Growth of Nb3Sn coating in tin vapor-diffusion process," *J. Vac. Sci. Technol. A* 37, 051509 (2019).
[59] Daniel Hall, "New Insights into the Limitations on the Efficiency and Achievable Gradients in Nb3Sn SRF Cavities", PhD thesis, Cornell University (2017).
[60] R.D. Porter et al., "Next Generation Nb3Sn Cavities for Linear Accelerators," Proc. LINAC'18, Beijing, China (2018).
[61] N. Sitaraman et al., "Ab Initio Study of Antisite Defects in Nb3Sn: Phase Diagram and Impact on Superconductivity," arXiv:1912.07576 [cond-mat.supr-con] (2019).
[62] S. Posen, N. Valles and M. Liepe, Radio frequency magnetic field limits of Nb and Nb3Sn, *Phys. Rev. Lett.* 115, 047001 (2015).
[63] R. Porter, Advancing the maximum accelerating gradient of niobium-3 tin superconducting radiofrequency accelerator cavities: RF measurements, dynamic temperature mapping, and material growth, Ph.D. thesis, Cornell University, Ithaca, NY, USA (2021).
[64] S. Posen, "Understanding and Overcoming Limitation Mechanisms in Nb3Sn Superconducting RF Cavities," Ph.D. thesis, Cornell University, Ithaca, NY, USA (2015).
[65] U. Pudasaini, G. Eremeev, C.E. Reece, H. Tian and M.J. Kelley, Electrochemical finishing treatment of Nb3Sn diffusion coated niobium, in Proceedings of the 18th International Conference on RF Superconductivity, Lanzhou, China, (2017).
[66] A. Gurevich, "Enhancement of rf breakdown field of superconductors by multilayer coating" *Appl. Phys. Lett.* 88, 012511 (2006).

[67] T. Kubo, Y. Iwashita, and T. Saeki, "Radio-frequency electromagnetic field and vortex penetration in multilayered superconductors" *Appl. Phys. Lett.* 104, 032603 (2014).

[68] V. Ngampruetikorn and J. Sauls, Effect of inhomogeneous surface disorder on the superheating field of superconducting RF cavities Vudtiwat Ngampruetikorn and J. A. Sauls, *Phys. Rev. Research* 1, 012015(R), (2019).

[69] G. Eremeev at al., "RF Performance Sensitivity to Tuning of $Nb_3Sn$ Coated CEBAF Cavities," Proc. of SRF'19, Dresden, Germany (2019).